\documentclass[prb,twocolumn,showpacs,showkeys,aps]{revtex4}  

\usepackage{graphicx}
\usepackage{amsmath}
\usepackage{amsfonts}
\usepackage{amssymb}

\date{\today }

\begin{document}

\title{Monte Carlo study of the transverse susceptibility in ordered arrays of magnetic nanoparticles}
\author{D. Kechrakos}
\email{dkehrakos@ims.demokritos.gr}
\author{K. N. Trohidou}
\affiliation{Institute of Materials Science "NCSR" Demokritos, GR-153 10, Athens, Greece}

\begin{abstract}
We present Monte Carlo simulations of the reversible transverse susceptibility (RTS) for a hexagonal array of dipolar interacting magnetic nanoparticles with random anisotropy. RTS curves with the bias-field in-plane and out-of-plane are compared.
With increasing temperature the RTS curves evolve from a three-peak ($H_C, \pm H_K$) structure to a double-peak and eventually a single-peak at the blocking temperature of the system. This trend is preserved for weak interactions.
Dipolar interactions at low temperature are responsible for the suppression of the $H_C$ peak in the out-of-plane geometry and its progressive merge to the $H_K$ peak with decreasing interparticle separation in the in-plane geometry. The $H_K$ peaks are located at higher field values in the out-of-plane geometry relative to the in-plane one.
When the bias field lies in-plane (out-of-plane) the $H_K$ peaks are shown to shift to lower (higher) field values with decreasing interparticle separation. The $H_C$ peak shifts to lower field values in both geometries. Our results are compared with recent experimental findings in self-assembled arrays of Fe nanoparticles.
\end{abstract}

\maketitle

\section{INTRODUCTION}

Ordered arrays of single-domain magnetic nanoparticles (NPs) produced by synthetic methods and self-assembly have attracted a lot of research effort over the last decade \cite{wil04}, motivated mainly by the wide range of potential technological applications that vary from high-density magnetic storage media \cite{sun00} to high-sensitivity field sensors and logic devices \cite{cow02}. Owing to the periodic arrangements of the NPs, their single-domain phase and their high monodispersity, they constitute ideal systems to gain basic understanding and possibly control of the role of interparticle magnetostatic interactions in their static and dynamical magnetic behavior. The issue of interparticle dipolar interactions has been addressed so far by a variety of experimental techniques including in most cases SQUID magnetometry and AC susceptibility measurements \cite{far05}, small-angle neutron scattering (SANS) \cite{iji05} and resonant magnetic X-ray scattering \cite{kor05} the last two methods being direct probes of magnetic correlations at the interparticle scale. These studies have provided ample evidence that the interplay between random anisotropy and dipolar interactions determine the magnetic behavior of the NP arrays. In particular observations such as, different values between the in-plane and normal-to-plane remanence magnetization \cite{rus00}, distribution of energy barriers with a larger width than the corresponding particle volume distribution \cite{woo01}, flat field-cooled magnetization curves \cite{pun01}, increase of the blocking temperature with increasing number of monolayers \cite{pod03,lui02} have been attributed to interparticle dipolar interactions. Model studies of the field and temperature dependent magnetization \cite{rus00,kec02} of NP arrays demonstrated the role of dipolar interactions and supported most of the experimental observations.

In addition to the above mentioned techniques, the reversible transverse susceptibility technique was theoretically  introduced by Aharoni \emph{et al}.\cite{aha57} According to the predictions of Aharoni \emph{et al}, under conditions of coherent rotation of the magnetization (Stoner-Wolfarth model) the dependence of RTS on bias field exhibits three distinct singularities, at the anisotropy fields $\pm H_{K}$ and at the coercive field $H_{c}$, which render the method suitable to probe the magnetic anisotropy. Almost thirty years later the first successful realization of the method \cite{par87} in BaFeO powders verified the theoretical predictions.
The RTS technique is anticipated to be particularly suitable for analysis of the anisotropy of single-domain magnetic NPs, as the magnetization dynamics of those is satisfactorily described by the coherent rotation assumption. However, there are various factors that make difficult the identification of the RTS peak positions and the extraction of the single-particle anisotropy strength. It has been previously demonstrated that particle size distribution rounds all three peaks\cite{hoa93}, orientational texture suppresses the coercivity peak \cite{hoa93}, interparticle interactions cause the coercive anisotropy peaks to merge \cite{cha94b,yan95}, and thermal relaxation of the moments causes large shifts of the coercive peak\cite{cha94a,yan95}.
Despite of these difficulties,the RTS technique was further developed and applied successfully to the analysis of the dynamical magnetic behavior of magnetic NP assemblies.\cite{spi00,spi01,spi02a,spi02b} These studies have demonstrated the capability of the technique to probe efficiently the transition of the NPs  from the blocked to the superparamagnetic regime by analyzing the field and temperature dependent RTS curves. Analysis of the peak structure, provided evidence of the presence of interparticle dipolar interactions in dense assemblies of $\gamma$ - Fe$_2$O$_3$ \cite{spi01} and Co  \cite{spi02a,spi02b} NPs. The issue of interparticle dipolar interactions and the induced collective magnetization dynamics was also studied in self-assembled arrays (SAA) of Fe \cite{pod03} and Fe-based \cite{pod05} NPs. The authors demonstrated the anisotropy between in-plane and out-of-plane RTS and suggested the existence of an intermediate phase between the blocked and superparamagnetic, where dipolar interactions dominate the dynamics of the assembly.

Previous theoretical investigations of the RTS in dipolar interacting NP assemblies have been performed within the mean-field approximation for the local field and a kinetic equation approach to the description of the thermal relaxation of the magnetic moments.\cite{yan95} This approach although it predicted correctly the fast decay of coercive peak with temperature it failed to give the expected shift of the anisotropy peak with temperature.
Micromagnetic studies of the transverse susceptibility implementing the Landau-Lifshitz-Gilbert (LLG) equations of motions for dipolar coupled magnetic moments\cite{spi03} have also appeared. The singularities of RTS in samples with uniaxial and cubic anisotropy were reproduced and the broadening of the peaks due to dipolar interactions was demonstrated. However thermal relaxation of the moments and orientational randomness of easy axes was not considered in that work and therefore the results cannot compare directly to existing experiments on SAA of magnetic NPs. Thermal fluctuations effects on the transverse susceptibility have also been considered implementing the stochastic LLG equations to calculate the imaginary part of RTS.\cite{vic04,cim06} Yuan and Victora \cite{vic04}studied granular films, where exchange interactions dominate. On the other hand, Cimpoesu \emph{et al}\cite{cim06} studied rectangular arrays of dipolar coupled NPs with perpendicular anisotropy and identified the effects of dipolar interactions on the signal of the imaginary part of RTS.
Despite the interesting conclusions demonstrated in the work of Cimpoesu \emph{et al} regarding the interplay of size distribution and dipolar effects, these can not be extended to the case of SAA of magnetic NPs, which are characterized by random anisotropy and a hexagonal arrangement of the NPs.

The aim of the present work is to model the behavior of the field dependent RTS of ordered arrays of magnetic NPs taking into account both thermal fluctuations and dipolar interaction effects. This is achieved implementing the Metropolis Monte Carlo (MC) algorithm. With this algorithm, the fluctuations in the the local field are treated exactly (up to the accuracy imposed by the calculation of the long-range dipolar forces) and  the magnetization correlations, required to describe collective magnetic behavior, are properly developed during the simulation.\cite{gon00} Finally, the arrangement of the NPs on the triangular lattice is a crucial ingredient of the model, because the dipolar interactions have a well established ferromagnetic (FM) character in this geometry.\cite{rus00,kec02,tro95}
In Section II we describe our model for the NP array and the simulation method for the calculation of RTS, in Section III we present numerical results for the in-plane and out-of plane RTS and finally in Section IV we discuss our results and make a connection to related experiments.

\section{ MODEL AND SIMULATION METHOD}

We consider N identical spherical particles with diameter $D$ forming a two-dimensional triangular lattice in the $xy$ plane with lattice constant $d$, where $d \geq D$. The size dispersion of the NPs can be neglected to a good approximation, since in most self-assembled samples a very narrow size distribution ($\sigma \lesssim 5\%$) is achieved.\cite{wil04} The particles are assumed single domain, with uniaxial anisotropy in a random direction, and they interact via dipolar forces. The total energy of the system is
\begin{eqnarray}
E = g\sum\frac{ \widehat{S}_{i}  \cdot \widehat{S}_{j}
 - 3(\widehat{S}_{i} \cdot \widehat{R}_{ij} )
    (\widehat{S}_{i} \cdot \widehat{R}_{ij} )  } {R_{ij}^{3}} \nonumber \\
 - k \sum ( \widehat{S}_{i}  \cdot \widehat{e_{i}} )^{2}
 - h \sum ( \widehat{S}_{i}  \cdot \widehat{H} )
\label{eq1}
\end{eqnarray}
where $\widehat{S}_{i}$ is the magnetic moment direction (spin) of particle i, $\widehat{e_{i}}$ is the easy-axis direction, and ${R}_{ij}$ is the center-to-center distance between particles i and j. Hats in Eq.~(\ref{eq1}) indicate unit vectors.
Three energy parameters enter Eq.~(\ref{eq1}), namely (i) the dipolar energy $g=m^{2} /d^{3}$, where $m =M_{s}V$ is the particle moment, $M_s$ the saturation magnetization density and $V$ the particle volume, (ii) the anisotropy energy $k=K_{1}V$, where $K_1$ is the uniaxial anisotropy energy density and (iii) the Zeeman energy $h=mH$, where $H$ is the applied dc field .
The relative strength of the energy parameters entering Eq.~(\ref{eq1}), the thermal energy $t=k_{B}T$, and the treatment history of the sample determine the micromagnetic configuration of the assembly. In all subsequent results we scale all energy parameters by the single particle anisotropy energy $(k=1)$. The transition from single-particle to collective behavior is determined solely by the ratio of the dipolar to the anisotropy energy $g/k=(\pi/6)(M_{s}^{2}/K_{1})(D/d)^{3}$. The reported values \cite{rus00,pun01,zha03} for fcc or hcp Co NPs are $g/k= 0.2-0.4(D/d)^{3}$, while for the soft $\epsilon$-Co phase, higher values are expected.\cite{pun01} For Fe NPs Farrell \emph{et al} \cite{far05} report $g/k= 1.54~(D/d)^{3}$ and Poddar \emph{et al} \cite{pod03} report $g/k= 2.8~(D/d)^{3}$.
To compare with the experiments of Poddar \emph{et al} we choose Fe NPs with $D=6.8nm$ at a separation $d \simeq 20nm$, which is an estimate from their TEM images. \cite{pod03} These values correspond to $g/k=0.1$ in our simulations.

\begin{figure}
\includegraphics[]{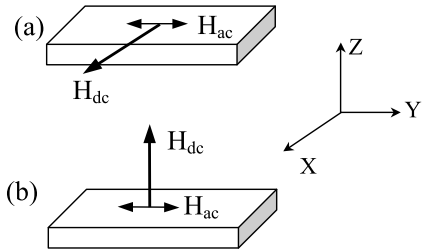}
\caption{
Sketch of the geometry used to calculate RTS. (a) In-plane geometry, used to obtain the parallel RTS ($\chi_T^\parallel$), and (b) out-of-plane geometry, used to obtain the perpendicular RTS ($\chi_T^\perp$). The NP superlattice is generated with basis vectors $\widehat{a}_{1}=(1,0)$ and $\widehat{a}_{2}=(\frac{1}{2},\frac{\sqrt{3}}{2})$.
}
\label{f1}
\end{figure}

Reversible transverse susceptibility measurements are performed with an ac measuring field ($H_{ac}$) perpendicular to the dc bias field. The ac field is weak ($\sim 10~Oe$) and its frequency lies in the rf regime ($f\sim10^{6}~Hz$).\cite{spi00,spi01,pod03}. These experimental conditions allow the following approximations to be adopted in the calculation of the RTS. First, since the amplitude of the measuring field is negligible compared to the saturation field ($\sim10^{3}~Oe$), the calculation of the susceptibility is performed in the zero-field limit $(H_{ac}=0)$. Second, the frequency dependence of the measuring field is neglected. This approximation is justified as long as the relaxation time of the NPs is large compared to the inverse frequency of the measuring field. Assuming that the NPs obey the N\'{e}el-Brown model for thermal relaxation, their relaxation time is given by \cite{cul72} $\tau=f_{0}^{-1}~exp(-KV/k_{B}T)$ with $f_{0}\simeq 10^{10} Hz$ and the blocking temperature $T_{b}=KV/ln(f_{0}\tau_{m})k_{B}$. For RTS measurements the characteristic measuring time is \cite{spi00,spi01} $\tau_{m}=2\cdot10^{-5}s$, and one obtains $\tau f= 100$ for $T = 0.94 T_{b}$ and $\tau f= 10$ for $T =1.16 T_{b}$.
Therefore, the static approximation is reasonably justified for temperatures up to $\sim 20\%$ above the blocking. In addition, our focus in the present work is on the static properties and their modification due to dipolar interactions, rather than on frequency dependent quantities, thus working in the static limit serves our purpose.
For in-plane measurements we take the bias field $H_{dc}$ along the $x$-axis and for out-of-plane measurements along the $z$-axis. In both cases the measuring field is assumed along the $y$-axis (see Fig.~\ref{f1}).The values of the RTS (per spin) are calculated from the fluctuations of the magnetization, $M_{y}=\sum S_{i}^{y}$, as

\begin{eqnarray}
\chi_T^{\parallel (\perp)}(H_{x(z)}) \equiv
\frac{1}{N} \left. {\frac{{\partial M_y }}{{\partial H_y^{ac} }}} \right|_{H_y^{ac}  = 0}
\nonumber\\
=\frac{1}{N k_{B}T} [ \langle M_{y}^{2} \rangle - \langle M_{y} \rangle^{2} ]
\label{eq2}
\end{eqnarray}
where $\chi_T^{\parallel (\perp)}$ is the in-plane (out-of-plane) RTS (see Fig.~\ref{f1}).
Simulations were performed for an ensemble of N=400 spins located in a simulation cell with dimensions $L_{x}=20d$ and $L_{y}=10\sqrt{3}d$ cut from a triangular lattice. For the interparticle interactions we used free boundaries along the $z$-axis and periodic boundaries in the $xy$ plane that diminish undesirable in-plane demagnetizing effects arising from free poles. The dipolar interactions were summed to infinite order in-plane, using the Ewald summation method for a quasi-two-dimensional system.\cite{grz00} For the simulation of the magnetic configuration under an applied field $H$ and at finite temperature $T$ we used the standard Metropolis MC algorithm with single-spin moves.\cite{bin98} According to this algorithm, a spin is chosen at random and it is rotated by a small angle. This is achieved by varying the Cartesian coordinates $(S_{i}^{x},S_{i}^{y},S_{i}^{z})$ of the moment randomly in the interval $(-\delta ,\delta )$, where $0<\delta \leq 1$, and renormalizing its new magnitude to unity. The new configuration is accepted with probability equal to $min[1,exp(-\Delta E/k_{B}T)]$, where $\Delta E$ is the change of the total energy. The value of $\delta $ is adjusted such that approximately $50\%$ of the attempted moves are accepted. Starting from a chosen spin configuration, the initial $10^{3}$ MC steps per spin (MCS) were used for relaxation of the system towards equilibrium and thermal averages were calculated over the subsequent $10^{4}$ MCS, allowing 10 MCS between sampling events to achieve statistical independence. The results were averaged over $N_{c}=100$ samples with different realizations of the random axes distribution and the thermal fluctuations. In all results presented below a sweep of the bias field from negative to positive values is performed with a step of $\Delta h=(1/30)k$. As a test of convergence a sweep in the opposite direction was performed in certain cases, but the deviations obtained in the positions of the peaks of RTS were well within the statistical errors.

\section{ NUMERICAL RESULTS}

\subsection{Remanence and coercivity}

\begin{figure}
\includegraphics[]{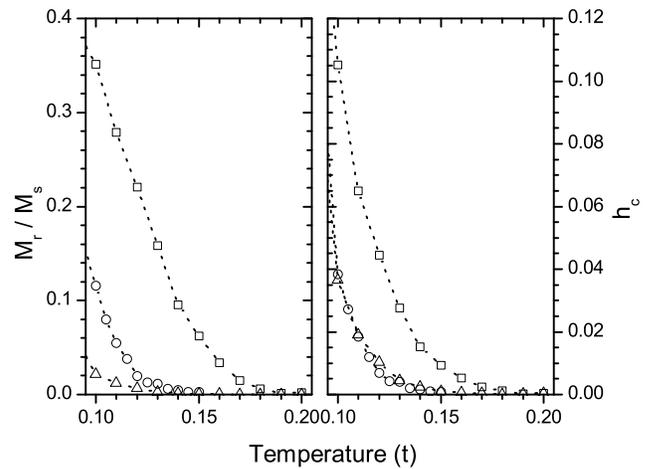}
\caption{Temperature dependence of remanence (left) and coercivity (right). Circles: non-interacting particles. Squares: interacting ($g=0.1$) particles and in-plane field. Triangles: interacting ($g=0.1$) particles and out-of-plane field.}
\label{f2}
\end{figure}

We discuss first the characteristics of the hysteresis loop for a NP array.
Dipolar interactions on a triangular lattice favor a ferromagnetic ground state and introduce an in-plane anisotropy.\cite{rus01,kec02,tro95}
The interaction-induced anisotropy produces a different magnetic behavior for in-plane and out-of-plane directions of the bias field.
In Fig.~\ref{f2} we show the temperature dependence of the remanence ($m_r=M_r/M_s$) and coercivity $(h_c)$ of a NP array with and without dipolar interactions.
We notice that the in-plane remanence ($m_r^\parallel$) is enhanced  while the out-of-plane remanence ($m_r^\perp$) is suppressed relative to the values for the non-interacting array.
Furthermore, the in-plane coercivity ($h_c^\parallel$) is clearly enhanced due to the extra barrier provided by the dipolar interactions, while the out-of-plane coercivity ($h_c^\perp$) shows a weaker dependence on the interaction strength.
For the non-interacting assembly the remanence and the coercivity exhibit vanishingly small values, $m_r(t)/m_r(0.05)\leq 1\%$ and $h_c(t)/h_c(0.05)\leq 0.5\% $, at temperatures above $t\simeq 0.14$, which is therefore the blocking temperature  for the non-interacting  assembly ($t_b^0$).\cite{comment01}
On the other hand, for the interacting assembly similarly small values, $m_r(t)/m_r(0.05)\leq 2\%$ and $h_c(t)/h_c(0.05)\leq 0.5\% $, are obtained above $t_b \simeq 0.17$.
At temperatures above $t_b^0$, the remanence and the coercivity are enhanced due to dipolar interactions.
This result defines an interesting temperature regime, $t_b^0 \leq t \leq t_b$, in which the thermal energy overwhelms the random anisotropy barrier and the dynamics of the NPs is governed by dipolar interactions.
The latter have a magnetizing effect at temperatures below $t_b^0$, as evidenced by the enhancement of $m_r^{\parallel}$. We refer to temperatures in the range $t_b^0 \leq t \leq t_b$  as the superferromagnetic (SFM) regime.
Notice also that the data presented in Fig.2 indicate that the SFM regime, is seriously suppressed in the out-of-plane geometry, due to the strong demagnetizing character of the dipolar interactions in this geometry.
Further down in this work we will discuss the behavior of the RTS in the SFM regime.

\subsection{Thermal evolution of transverse susceptibility}

\begin{figure}
\includegraphics[]{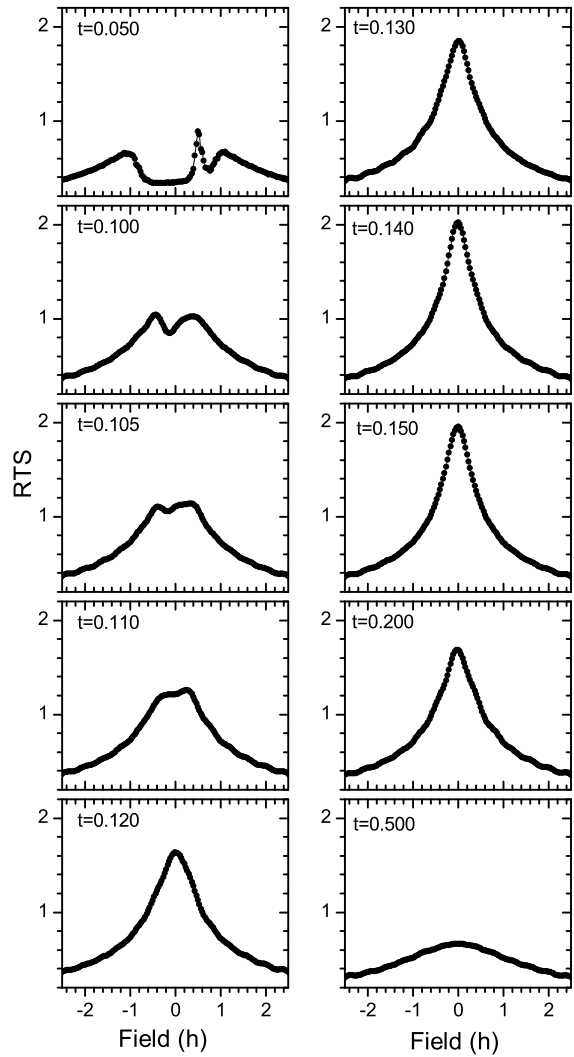}
\caption{Evolution of RTS with temperature for a non-interacting ($g=0$) array. The magnetic field is swept from negative to positive values.}
\label{f3}
\end{figure}

The field-dependent RTS of a non-interacting assembly is shown in Fig.~\ref{f3} for various temperatures that extend from low ($t=0.05$) to high values ($t=0.50$).
At low temperature, the RTS curve shows two broad peaks at symmetric positions that correspond to the anisotropy field $\pm h_K$ and a third sharper peak at the coercive field $h_c$.
The presence of three peaks in the low-temperature RTS curve is in accordance with the theoretical predictions of Aharoni \emph{et al}\cite{aha57} and previous numerical studies.\cite{cha94a,hoa93}
Notice however that the anisotropy peaks appear at $h_K\simeq\pm 1.0$ which is lower than the zero-temperature value $h_K= \pm 2$, because of the non-zero temperature at which the calculation is performed.
Indeed, for an isolated NP, the anisotropy peak is centered at the field required for an irreversible switch of the particle's moment.
Therefore, the thermal energy assists the switch and causes a shift of the anisotropy peak towards lower field values.
In previous numerical calculations of RTS at finite temperature \cite{cha94a}, the authors implemented a rate equation approach and found that the positions of the anisotropy peaks are insensitive to the temperature.
This result does not agree with RTS measurements in NP assemblies that show a clear downshift of the anisotropy peaks with temperature \cite{spi00,spi01}.
Our approach, that goes beyond the mean field description of the thermal fluctuations of the magnetization predicts the expected downshift of the anisotropy peaks with temperature.
As shown in Fig.~\ref{f2}, the coercivity decays fast with temperature and in Fig.~\ref{f3} we show that already for temperatures $t=0.10-0.11$, which lie well below the blocking ($t_b^0=0.14$) the coercivity peak of RTS is suppressed, leading to a pair of asymmetric anisotropy peaks.
The broader among the two anisotropy peaks appears at a positive field, namely at a strong enough field to cause reversal of the magnetization.
The formation of a broad anisotropy peak due to its merge with the coercivity peak has also been observed in RTS measurements in Fe-based \cite{spi01} and Co  \cite{spi02a,spi02b} NP assemblies.
The fast downshift of the $h_c$ peak relative to the slow shift of the $h_K$ peaks was also found in previous numerical studies of RTS and suggests that extraction of the $h_c$ field from RTS measurements is much more difficult than of the $h_K$ field, even for monodisperse and extremely dilute samples. \cite{cha94a,yan95}
When the temperature is increased above $t=0.11$ (Fig~\ref{f3}), the anisotropy peaks rise and merge to a single narrow peak that subsequently gets broader and lower above $t = 0.14$.
The susceptibility at zero field (Fig.~\ref{f3}) exhibits its maximum value at $t=0.14$, which is the blocking temperature of the assembly.
Above the blocking , the zero-field RTS drops with temperature according to the Curie law ($\chi \sim 1/T$).
In summary the thermal evolution of the RTS curves, shown in Fig.~\ref{f3} is characterized by a three-peak mode at very low temperature followed by to a double-peak mode due to the merge of the coercivity and anisotropy peaks at intermediate temperature, and finally, a single-peak mode above the blocking temperature ($t > t_b^0$).

\begin{figure}
\includegraphics[]{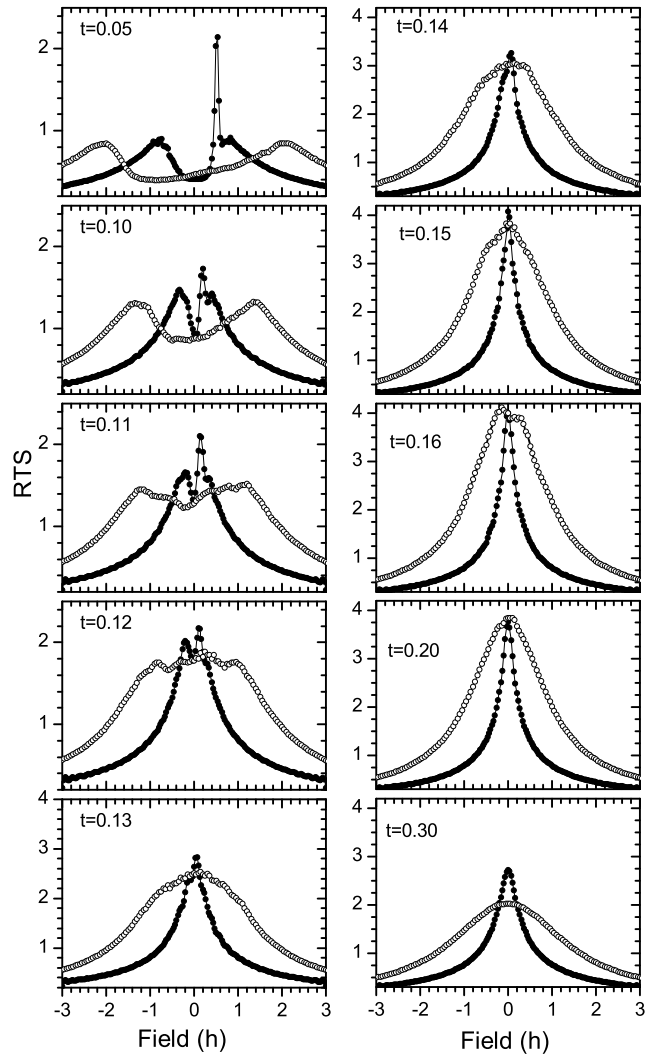}
\caption{Evolution of RTS with temperature for a dipolar interacting ($g=0.1$) array. The magnetic field is swept from negative to positive values. Closed circles : in-plane field. Open circles : out-of-plane field.}
\label{f4}
\end{figure}

When weak interparticle interactions ($g=0.1$) are switched on, they do not modify the overall trend of the RTS curves with temperature, namely the transition from three-peaks to a double-peak and finally to a single-peak (Fig.~\ref{f4}).
However, even weak interactions introduce noticeable differences between the in-plane susceptibility ($\chi_T^{\parallel}$) and the out-of-plane one ($\chi_T^{\perp}$).
Starting from the low-temperature regime ($t=0.05$) we notice that the $h_c $ peak is suppressed in the $\chi_T^{\perp}$ curve while it is quite pronounced in the $\chi_T^{\parallel}$ curve. The in-plane anisotropy induced by dipolar interactions transforms the spherical distribution of the easy axes directions to a quasi two-dimensional distribution that causes suppression of the $h_c^\perp$ peak.
A similar suppression of the $h_c$ peak was demonstrated by Hoare \emph{et al}\cite{hoa93} due to easy axes texture in a non-interacting assembly. For dipolar coupled arrays, the interparticle interactions produce a similar effect to texturing via the long range demagnetizing field that forces the moments to stay in-plane.
At all temperatures the $\chi_T^{\perp}$ curve is broader than the $\chi_T^{\parallel}$ curve, because of the larger saturation field in the out-of-plane geometry. The slow saturation arises due to the in-plane anisotropy induced by dipolar interactions which renders magnetically harder the z-axis relative to the x,y-axes.
A further difference occurring between the two measuring geometries at low temperature is that the anisotropy peaks are located at higher fields in the case of out-of-plane geometry ($h_K^\perp > h_K^\parallel $). This feature is in agreement with experimental observations in Fe NP arrays.\cite{pod03}

\begin{figure}
\includegraphics[]{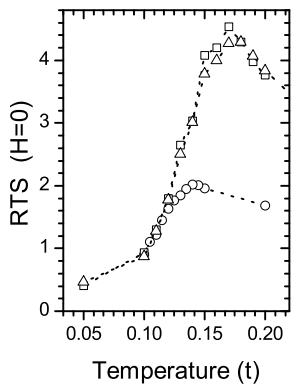}
\caption{Temperature dependence of the zero-bias RTS. Circles: $g=0$. Squares: $g=0.1$ and in-plane geometry. Triangles: $g=0.1$ and out-of-plane geometry.}
\label{f5}
\end{figure}

A physical argument to justify this behavior is the following : The position of the anisotropy peak corresponds the field required for an irreversible switch of the particle's moment over the uniaxial anisotropy barrier. Dipolar interactions induce an  easy-plane anisotropy additionally to the uniaxial one. Therefore, the in-plane switching is facilitated by the presence of more than one easy directions and consequently the switching field is reduced, leading to a downshift of the corresponding RTS peak. On the contrary, out-of-plane switching is obstructed by the additional easy-plane that tends to keep the moment inside the xy-plane and the out-of-plane switching field is enhanced.
Eventually, as the temperature rises thermal fluctuations overwhelm the dipolar anisotropy and the difference between the $\chi_T^{\parallel}$ and $\chi_T^{\perp}$ curves is reduced, as shown in Fig.~\ref{f4} for $t=0.30$.

To extract the blocking temperature from the susceptibility curves, we plot in Fig.~\ref{f5} the zero-field susceptibility $\chi_T(H=0)$ as a function of temperature.
Provided that the peak of $\chi_T(T;H=0)$ occurs at the blocking temperature of the system,  we obtain from Fig.~\ref{f5} the values $t_b^0 \simeq 0.14$ and $t_b \simeq 0.17$. These values are in satisfactory agreement with the values obtained earlier from the temperature dependence of the remanence and coercivity (Fig.~\ref{f2}).
Notice that in contrast to the experimental situation, where magnetization and RTS measurements have different measuring times ($\tau_{SQUID}\sim 10^2 s, \tau_{RTS}\sim 10^{-5} s$) \cite{spi02b,pod03} leading to different estimates of the blocking temperature, in our case both magnetization and susceptibility data are obtained during the same Monte Carlo relaxation process and therefore correspond to measurements at the same time scale.

\begin{figure}
\includegraphics[]{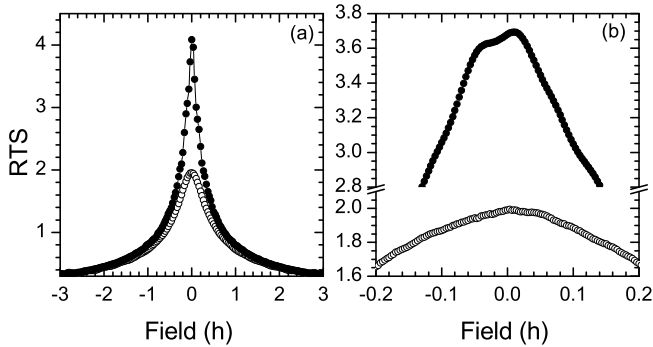}
\caption{Field dependent RTS in the superferromagnetic regime ($t=0.15 $). Open circles : $g=0$. Closed circles : $g=0.1$ and in-plane geometry. Field step : (a) $\Delta h=(1/30)k$ and (b) $\Delta h=(1/300)k$.}
\label{f6}
\end{figure}

A point concerning the thermal evolution of the RTS curves across the blocked to superparamagnetic regime is next.
The issue is whether the transition from the two-peak to the single peak mode occurs \emph{below} or \emph{at} the blocking temperature.
Previous calculations for non-interacting assemblies\cite{spi01} point to the second scenario, namely that the double peak structure is preserved right up to the blocking temperature.
However, in recent experiments in densely packed Fe NP arrays Poddar \emph{et al} \cite{pod03} argued that the anisotropy peaks merge at a temperature ($T_{cross}$) that lies \emph{below} the blocking temperature of the array.
They supported this observation by arguing that the merge of the peaks indicates the overcome of the uniaxial anisotropy barrier by the thermal energy, while the maximization of the RTS peak corresponds to the overcome of the interaction-induced barrier by the thermal energy.

\begin{figure}
\includegraphics[]{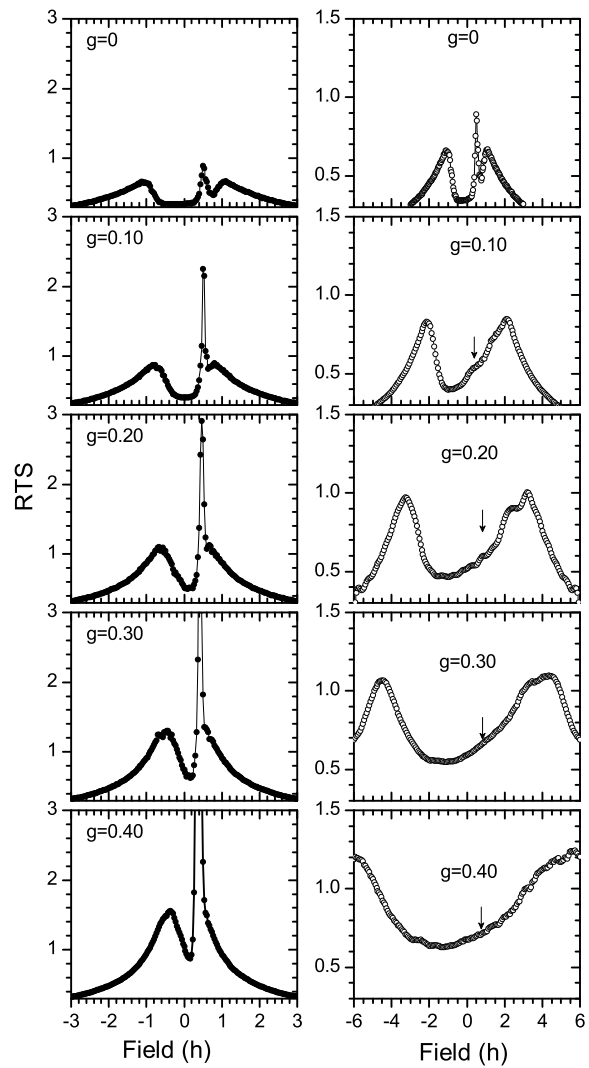}
\caption{Low-temperature ($t=0.05$) RTS for variable dipolar strength. Closed circles : in-plane geometry. Open circles : out-of-plane geometry. The arrows in the out-of-plane data indicate the position of the coercive field.}
\label{f7}
\end{figure}

We examined this argument by performing more refined simulations in the field range around the peak. In particular, we have performed a high-resolution simulation at a temperature $t=0.15$, that lies in the SFM regime (see Fig.~\ref{f5}), using an order of magnitude smaller field step $dh=(1/300)k$.
The choice of temperature is a typical value in the SFM regime, in which, as discussed earlier, thermal fluctuations are adequate to overcome the uniaxial anisotropy but the system remains ferromagnetic due to dipolar interactions.
The results are shown in Fig.~\ref{f6}, where it is demonstrated that the double peak structure for the interacting array is preserved at this temperature, while the non-interacting array shows a perfectly broad single peak. Repeating the simulations at higher temperatures but still below the blocking temperature $(t<0.17)$ we reached a double-peak curve with a closer distance between the peaks and reduced asymmetry. Therefore our simulations show that in the SFM regime ($t_b^0 \leq t \leq t_b$) the double peak structure of the RTS curve is preserved and it transforms to a single peak \emph{at} the blocking temperature, similarly to the case of non-interacting assemblies.

\subsection{Evolution of transverse susceptibility with particle spacing}

\begin{figure}
\includegraphics[]{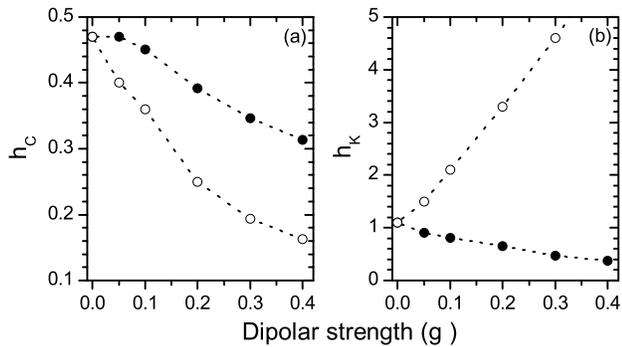}
\caption{Variation of the (a) coercivity peak and (b) anisotropy peak position with the dipolar strength at low temperature ($t=0.05$). Closed circles : in-plane geometry. Open circles : out-of-plane geometry.}
\label{f8}
\end{figure}

Varying the interparticle spacing ($d$), modifies the strength of the dipolar coupling as $g\sim 1/d^3$. We show in Fig.~\ref{f7} the in-plane and out-of-plane susceptibilities at low temperature ($t=0.05$) for increasing dipolar strength.
The most striking difference between the two geometries is that as the interaction strength increases, the anisotropy peaks of $\chi_T^{\parallel}$ shift to lower fields while those of $\chi_T^{\perp}$ shift to higher values.
The peak positions for the two geometries are plotted in Fig~\ref{f8} as a function of the dipolar strength.

The dependence of the anisotropy peaks on the coupling strength has been discussed above and is attributed to the easy-plane anisotropy induced by the dipolar interactions. A linear decrease of the anisotropy field with packing density has been previously reported in randomly packed magnetic NPs \cite{sol97}. Our simulations indicate that the same behavior is expected in ordered arrays and the in-plane geometry.
The similarity between these two situations lies in the fact that in both cases the interaction-induced anisotropy reduces the barrier for an irreversible switching of the moments. On the contrary, in the out-of-plane geometry the anisotropy peak shifts to higher fields with increasing coupling strength since dipolar interactions inhibit the irreversible switching along the z-axis. The outcome of the interactions for the ordered array is to produce higher values for the out-of-plane anisotropy field than the in-plane one ($ h_k^{\perp} > h_k^{\parallel} $), which is in agreement with the RTS measurements \cite{pod03}in arrays of Fe NPs.

\section{CONCLUSIONS}

We studied the reversible transverse susceptibility in hexagonal arrays of anisotropic magnetic nanoparticles using Monte Carlo simulations, in order to investigate the effects of interparticle dipolar interactions.
We found that below the blocking temperature dipolar interactions are responsible for a series of characteristic features in the RTS curves. Namely, (i) the suppression of the $H_{C}$ peak of $\chi_T^{\perp}$, (ii) the location of $\pm H_{K}$ peak of $\chi_T^{\perp}$ at higher fields than the corresponding peaks of $\chi_T^{\parallel}$, (iii) the downshift (upshift) of the $H_K^{\parallel}$ ($H_K^{\perp}$) peak with increasing dipolar strength or equivalently, decreasing interparticle spacing, (iv) the slower saturation with bias field of  $\chi_T^{\perp}$ relative to $\chi_T^{\parallel}$.
These results are in agreement with recent measurements of in-plane and out-of-plane RTS  in ordered arrays of Fe nanoparticles.\cite{pod03}
With respect to the thermal evolution of the RTS curves, we showed that with increasing temperature both the $H_{K}$ and $H_{C}$ peaks shift to lower field values and merge to a single peak leading to a double peak structure. Dipolar interactions are shown (Fig.~\ref{f4}) to make the $H_C^{\parallel}$ peak more persistent to thermal fluctuations.
At the blocking temperature of the system the anisotropy peaks merge and the zero-field RTS is maximized. The same thermal evolution of the peaks is followed in weakly interacting arrays and agrees with the experimental observations in Fe NP arrays.\cite{pod03}
Therefore, our simulations support the capability of RTS measurements to provide useful information related to the dynamical state of interacting nanoparticle assemblies.
A minor difference between our simulations and measurements on SAA of Fe nanoparticles,\cite{pod03} is that experimentally the transition from a double-peak to a single-peak structure of RTS was observed at a temperature $T_{cross} < T_b$ and attributed to a transition to a superferromagnetic regime. Our simulations showed that the double peak structure persists up to $T_b$.
Possible reasons for this discrepancy should be sought for in the limitations imposed by the instrumental resolution or in the static approximations adopted in our model.  This point requires further investigation.

\begin{acknowledgments}
Work supported by EC project NANOSPIN (Contract No NMP4-CT-2004-013545)
\end{acknowledgments}


\begin{thebibliography}{99}
\bibitem{comment01} Our simulations predict a higher blocking temperature $(t_b/k \sim 1/7.7)$ than a typical magnetometry experiment $(t_b/k \sim 1/27.6)$ or an RTS-rf experiment $(t_b/k \sim 1/11.5)$ owing to the number of MCS used. Prohibitively long simulations would be required in order to get values of $t_b$ close to the experimental ones, while the trend of the results when varying the physical parametes invlolved (for example, g, h) is not expected to be affected.
\end{thebibliography}
\end{document}